\documentclass[twocolumn,nofootinbib,amsmath,amssymb,a4paper]{revtex4}

\usepackage{graphicx}% Include figure files
\usepackage{dcolumn}% Align table columns on decimal point
\usepackage{bm}

\begin{document}

\title{Recent Developments in the PQCD Approach~\footnote{Talk presented
at the 4th International Workshop on the CKM Unitarity Triangle
(CKM2006), Nagoya, Japan, Dec.~12-16, 2006.}
}% Force line breaks with \\

\author{Satoshi Mishima}
 \email{mishima@ias.edu}
\affiliation{%
School of Natural Sciences, Institute for Advanced Study, 
Princeton, NJ 08540, U.S.A.
}%

\begin{abstract}
We review recent developments in the perturbative QCD approach to
 exclusive hadronic $B$ meson decays. We discuss the important
 next-to-leading-order corrections to $B\to \pi K$, $\pi\pi$, and the
 penguin-dominated $B\to PV$ modes, where $P$ ($V$) is a pseudo-scalar
 (vector) meson. 
\end{abstract}

\maketitle

\section{Introduction}

$B$ factory experiments have accumulated a lot of data on exclusive
hadronic $B$ decays, and have reported many interesting
results~\cite{HFAG}. Some observables, mixing-induced CP asymmetries for
$b\to s$ penguin modes, branching ratios and direct CP asymmetries for
$B\to \pi K$ and $\pi\pi$, and other observables, have exhibited some
deviations from na\"{\i}ve expectations in the Standard Model (SM). 
It is necessary to go beyond na\"{\i}ve estimations for understanding
the observed deviations, by including subdominant contributions, such as
spectator and annihilation diagrams, and higher-order corrections. Most
of the calculations of $B$ decay amplitudes rely on the factorization of
decay amplitudes into a product of short-distance and long-distance
physics.  QCD-improved factorization (QCDF)~\cite{BBNS} and
soft-collinear effective theory (SCET)~\cite{SCET} are based on
collinear factorization theorem, but the perturbative QCD (PQCD)
approach~\cite{PQCD} is based on $k_T$ factorization theorem.  
Employing collinear factorization theorem, some decay amplitudes involve
a singularity arising from the end-point region of parton momentum
fractions.  An end-point singularity implies that the decay amplitude is
dominated by soft dynamics and cannot be further factorized. Such soft
contributions are regarded as phenomenological parameters, which are
fitted from experimental data.

\begin{figure}
\includegraphics{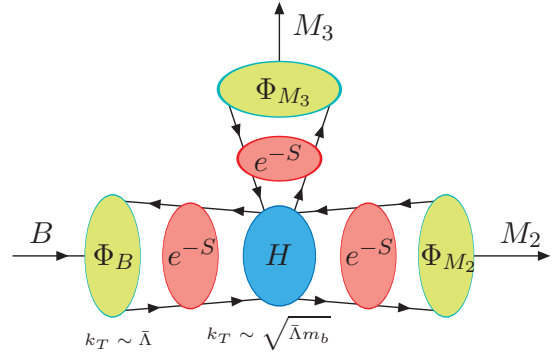}
\caption{\label{fig:ktfact} PQCD factorization theorem for $B\to M_2M_3$.}
\end{figure}

In the PQCD approach with $k_T$ factorization theorem, there is no
end-point singularity because of the Sudakov
factor~\cite{Li:1992nu}. All amplitudes then can be factorized into
parton distribution amplitudes $\Phi$, the Sudakov factors $e^{-S}$, the
jet function $J$, and the hard kernel $H$ as~\cite{CL} 
\begin{equation}
A(B\to M_2 M_3)
=
\Phi_{M_2} \otimes \Phi_{M_3} \otimes 
H \otimes J \otimes e^{-S} \otimes 
\Phi_B 
\;,
\end{equation}
%
%where $\otimes$ stands for convolutions in both longitudinal and
where the symbols $\otimes$ stand for convolutions in both longitudinal and
transverse momenta of partons. 
The distribution amplitudes, which are universal in the processes under
consideration, are determined from experiments, the light-cone QCD sum
rules, lattice calculations, or other non-perturbative methods, and are
the main source of uncertainty. 
The hard kernel is characterized by the hard scale 
$t\sim{\cal O}(\sqrt{\bar\Lambda m_b})$, where $\bar\Lambda$ is a
hadronic scale and $m_b$ the $b$ quark mass, and can be evaluated as an
expansion in powers of $\alpha_s(t)$ and $\bar\Lambda/t$. 
The schematic picture of the factorization theorem is displayed in
Fig.~\ref{fig:ktfact}. 
The parton transverse momenta $k_T $ in the mesons are of the order of
$\bar\Lambda$. With infinitely many collinear gluon emissions, $k_T$
accumulate and reach ${\cal O}(\sqrt{\bar\Lambda m_b})$ in the hard
kernel. It ensures the absence of the end-point
singularities~\cite{Li:1992nu}. 
The hard kernels of the spectator and the annihilation contributions, as
well as the emission contribution, are calculable and start from 
$O(\alpha_s)$. PQCD has applied to various two-body $B$ decays at
leading order (LO) in $\alpha_s$, and has made reasonable predictions
for many modes. Note that PQCD predicts a large direct CP asymmetry in
the $B^0\to\pi^\mp K^\pm$ mode as a result of a large strong phase
arising from scalar-penguin annihilation diagrams~\cite{PQCD}.  
Recently, a part of next-to-leading-order (NLO) contributions have been
included in the PQCD approach~\cite{Li:2005kt,Li:2006cv,Li:2006jv}. 
\begin{figure*}
\includegraphics{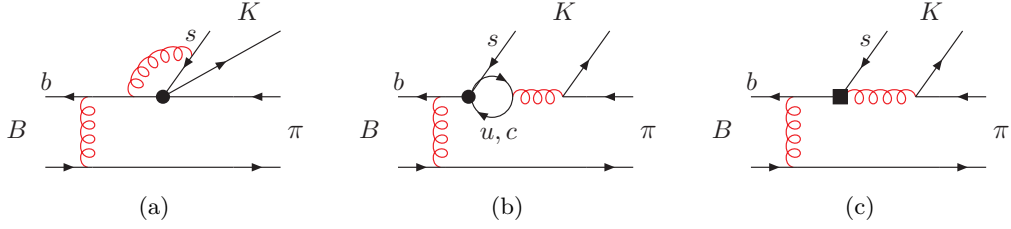}
\caption{\label{fig:nlo} Examples of NLO diagrams: (a) the vertex
 correction, (b) the quark loop, and (c) the magnetic penguin, where the
 dots and the square represent local operators.}
\end{figure*}
Only the most important NLO contributions coming from the NLO Wilson
coefficients, the vertex corrections, the quark loops, and the magnetic
penguin, shown in Fig.~\ref{fig:nlo}, 
have been considered~\cite{Li:2005kt}, because other NLO corrections,
mainly to the $B$ meson transition form factors, can be eliminated by
choosing an appropriate scale 
$t\sim {\cal O}(\sqrt{\bar\Lambda m_b}\,)$~\cite{Brodsky:1982gc}. 
In this talk, we summarize NLO PQCD predictions for $B\to \pi K$,
$\pi\pi$, and the penguin-dominated $B\to PV$ modes, $P (V)$ being a
pseudo-scalar (vector) meson.

This talk is organized as follows: In Sec.~\ref{sec:pik}, we discuss the
branching ratios and the CP asymmetries of $B\to\pi K$ and $\pi\pi$ with
the NLO corrections. The predictions of the penguin-dominated $B\to PV$
decays are presented in Sec.~\ref{sec:pv}. The mixing-induced CP
asymmetries for $b\to s$ penguin modes are in
Sec.~\ref{sec:mixing}. Section~\ref{sec:summary} is a summary.

\section{$B\to\pi K$ Puzzle \label{sec:pik}}

The current data of the direct CP asymmetries of $B\to\pi K$ and the
branching ratios of $B\to\pi\pi$ listed in Table~\ref{tab:pik_pipi} are
inconsistent with the na\"{\i}ve expectations 
\begin{eqnarray}
A_{\rm CP}(B^\pm\to \pi^0 K^\pm) &\approx& A_{\rm CP}(B^0\to \pi^\mp K^\pm)
\;,\nonumber\\
{\rm Br}(B^0\to\pi^\pm\pi^\mp) &\gg& {\rm Br}(B^0\to\pi^0\pi^0)
\;. 
\label{eq:rel1}
\end{eqnarray} 
\begin{table}
\caption{\label{tab:pik_pipi} Branching ratios in units of $10^{-6}$ and
 direct CP asymmetries in percentage for the $B\to\pi K$ and $\pi\pi$
 decays~\cite{Li:2005kt}.} 
\begin{ruledtabular}
\begin{tabular}{c|ccc|ccc}
& \multicolumn{3}{c|}{Br($10^{-6}$)}
& \multicolumn{3}{c}{$A_{\rm CP}(10^{-2})$}
\\
Mode & Data~\cite{HFAG} & LO & NLO & Data~\cite{HFAG} & LO & NLO
\\
\hline
$\pi^\pm K^0$ &
 $23.1 \pm 1.0$ &
 $17.0$ &
 $23.6^{+14.5}_{-8.4}$ &
 $0.9 \pm 2.5$ & 
 $-1$ &
 $0\pm 0$
\\
$\pi^0 K^\pm$ &
 $12.8 \pm 0.6$ &
 $10.2$ &
 $13.6^{+10.3}_{-5.7}$ &
 $4.7 \pm 2.6$ & 
 $-8$ &
 $-1^{+3}_{-6}$
\\
$\pi^\mp K^\pm$ & 
 $19.7 \pm 0.6$ &
 $14.2$ &
 $20.4^{+16.1}_{-8.4}$ &
 $-9.3 \pm 1.5$ & 
 $-12$ &
 $-10^{+7}_{-8}$ 
\\
$\pi^0 K^0$ & 
 $10.0 \pm 0.6$ &
 $5.7$ &
 $8.7^{+6.0}_{-3.4}$ &
 $-12\pm 11$ &
 $-2$ &
 $-7^{+3}_{-4}$
\\
\hline
$\pi^\mp\pi^\pm$ & 
 $5.2 \pm 0.2$ &
 $7.0$ &
 $6.5^{+6.7}_{-3.8}$ &
 $39 \pm 7$ & 
 $14$ &
 $18^{+20}_{-12}$
\\
$\pi^\pm\pi^0$ &
 $5.7 \pm 0.4$ &
 $3.5$ &
 $4.0^{+3.4}_{-1.9}$ &
 $4 \pm 5$ &
 $0$ &
 $0\pm 0$
\\
$\pi^0\pi^0$ &
 $1.31 \pm 0.21$ &
 $0.12$ &
 $0.29^{+0.50}_{-0.20}$ &
 $36^{+33}_{-31}$ &
 $-4$ &
 $63^{+35}_{-34}$
\\
\end{tabular}
\end{ruledtabular}
\end{table}
These relations can be understood in the topological-amplitude
decompositions (see Ref.~\cite{Li:2005kt} and references therein):  
\begin{eqnarray}
%A(B^+\to\pi^+ K^0)&=&P'
%\;,\nonumber\\
\sqrt{2}A(B^+\to \pi^0 K^+)&=&
-P'-P'_{ew}
-\left(T'+C' \right)e^{i\phi_3}
\;,\nonumber\\
A(B^0\to \pi^- K^+)&=&-P'-T'e^{i\phi_3}
\;,\nonumber\\
%\sqrt{2}A(B^0\to \pi^0K^0)&=&
%P' - P'_{ew}
%-C'e^{i\phi_3}
%\;,\nonumber\\
A(B^0\to \pi^+\pi^-)&=&-T - Pe^{i\phi_2}
\;,\nonumber\\
%\sqrt{2}A(B^+\to \pi^+\pi^0)&=&
%-T - C
%-P_{ew}e^{i\phi_2}
%\;,\nonumber\\
\sqrt{2}A(B^0\to \pi^0\pi^0)&=&
\left(P-P_{ew}\right)
e^{i\phi_2}-C
\;,
\label{eq:pik_amp}
\end{eqnarray}
where $T^{(\prime)}$, $C^{(\prime)}$, $P^{(\prime)}$, and
$P_{ew}^{(\prime)}$ stand for the color-allowed tree, color-suppressed
tree, penguin, and electroweak penguin amplitudes, respectively, and
$\phi_2$ and $\phi_3$ are the Cabbibo-Kobayashi-Maskawa (CKM) phase
defined by 
$V_{ub}=|V_{ub}|\exp(-i\phi_3)$, $V_{td}=|V_{td}|\exp(-i\phi_1)$, and
$\phi_2=180^\circ - \phi_1 - \phi_3$.  
Assuming the hierarchies, $P' > T',\, P'_{ew} > C'$ for $B\to\pi K$ and 
$T > C,\, P > P_{ew}$ for $B\to\pi\pi$, the relations in
Eq.~(\ref{eq:rel1}) can be derived, where the asymmetries are written as 
\begin{eqnarray}
A_{\rm CP}(B^\pm\to\pi^0 K^\pm)
&\simeq&
2\, {\rm Im}\left[\frac{T'+C'}{P'+P'_{ew}}\right]\sin\phi_3
\nonumber\;,\\
A_{\rm CP}(B^0\to \pi^\mp K^\pm)
&\simeq& 
2\,{\rm Im}\left[\frac{T'}{P'}\right]\sin\phi_3
\;. 
\end{eqnarray}
The mixing-induced CP asymmetry for $B^0\to\pi^0 K_S$ has exhibited
another puzzle. Hence, the current data seem to require a large
$C^{(\prime)}$, a large $P^{(\prime)}_{ew}$ with a new CP violating
phase, or both of them (see, {\it e.g.}, Refs.~\cite{Li:2005kt,PUZZLE}
and references therein).

\begin{figure}
\includegraphics{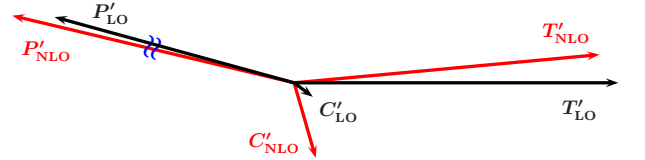}
\caption{\label{fig:topology} Tree $T'$, color-suppressed tree $C'$, and
 penguin $P'$ amplitudes in the complex plane, where $|P'|$ is much
 larger than $|T'|$ and $|C'|$~\cite{Li:2005kt}. The amplitudes with the 
 subscript LO (NLO) were calculated up to LO (NLO) in $\alpha_s$.}
\end{figure}

The LO PQCD predictions follow the na\"{\i}ve expectations as listed in
the third and sixth columns of Table~\ref{tab:pik_pipi}~\cite{PQCD}. 
The effect of the NLO corrections was studied in Ref.~\cite{Li:2005kt}.  
The sum of the quark loops and the magnetic penguin reduces the penguin
amplitudes by about 10\% in the $B\to\pi K$ decays and affects the CP
asymmetries little. 
The vertex corrections much affect $C^{\prime}$, associated with the
effective Wilson coefficient 
\begin{equation}
a_2(\mu)\ =\ 
\left( C_1(\mu) + \frac{C_2(\mu)}{N_c} \right)
+ \frac{\alpha_s(\mu)}{9\pi}\,V_2(\mu)\, C_2(\mu)
\;, 
\end{equation}
where $N_c=3$ is the number of colors and $V_2$ the loop function from
the vertex corrections.
The vertex correction term has a large coefficient $C_2$, whereas the LO
ones cancel between $C_1$ and $C_2/N_c$. 
Therefore the vertex corrections enhance $C'$ significantly and rotate
its phase, such that $T'+C'$ is anti-parallel to $P'$ as shown in
Fig.~\ref{fig:topology}, where $C'$ is still subdominant compared with
$T'$. 
Thus, $A_{\rm CP}(B^\pm\to\pi^0 K^\pm)$ vanishes and NLO PQCD
predicts the pattern~\cite{Li:2005kt}  
\begin{equation}
|A_{\rm CP}(B^\pm\to\pi^0 K^\pm)|
\ \ll\  
|A_{\rm CP}(B^0\to\pi^\mp K^\pm)|
\;,
\label{eq:pik}
\end{equation}
as listed in the last column of Table~\ref{tab:pik_pipi}.

Similarly, $C$ for $B\to\pi\pi$ is enhanced by the vertex corrections,
but it is insufficient to accommodate Br$(B^0\to\pi^0\pi^0)$ to the
measured value~\cite{Li:2005kt}. NLO PQCD predicts $|C/T|\approx 0.2$
for $B\to\pi\pi$, though a much larger $|C/T|\approx 0.8$ is required to
explain the observed Br$(B^0\to\pi^0\pi^0)$. 
Note that the larger $C$ also contributes to the $B^0\to\rho^0\rho^0$
mode. The NLO PQCD prediction for Br$(B^0\to\rho^0\rho^0)$ has almost
reached the experimental upper bound~\cite{Li:2006cv}. 
Thus it is unlikely to accommodate both Br$(B^0\to\pi^0\pi^0)$ 
and Br$(B^0\to\rho^0\rho^0)$ to the data simultaneously.

\section{Penguin-dominated $B\to PV$ Decays \label{sec:pv}} 

It is expected that the NLO corrections also affect penguin dominated
$B\to PV$ modes, such as $B\to \pi K^*$, $\rho K$, $\omega K$, and 
$\phi K$. In QCDF, those branching ratios are usually smaller than the
observed values~\cite{BBNS}. On the other hand, in PQCD, the predicted
branching ratios for $B\to \phi K$ are large enough to explain the 
data~\cite{Mishima:2001ms}.

\begin{figure*}
\includegraphics{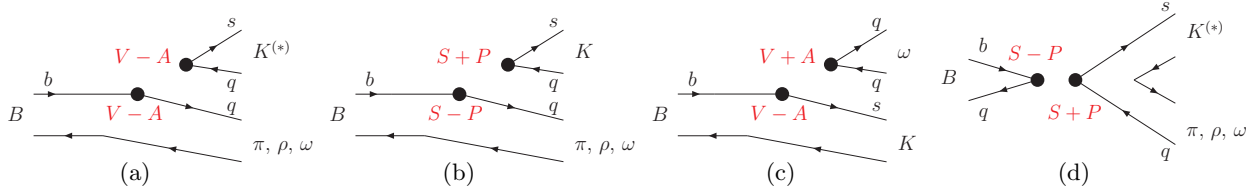}
\caption{\label{fig:bpv} Penguin emission and annihilation amplitudes,
 where the emission ones (a), (b), and (c) are associated with 
 $a_4$, $a_6$, and $a_5$, respectively.}
\end{figure*}

The LO and NLO PQCD predictions for the $B\to PV$ branching
ratios~\cite{Li:2006jv}, as well as the current data~\cite{HFAG}, are
summarized in Table~\ref{tab:pv}. 
\begin{table}
\caption{\label{tab:pv} Branching ratios in units of $10^{-6}$ and
 direct CP asymmetries in percentage for the $B\to\pi K^*$, $\rho K$,
 $\omega K$, and $\phi K$ decays~\cite{Li:2006jv}.}
\begin{ruledtabular}
\begin{tabular}{c|ccc|ccc}
& \multicolumn{3}{c|}{Br($10^{-6}$)}
& \multicolumn{3}{c}{$A_{\rm CP}(10^{-2})$}
\\
Mode & 
Data~\cite{HFAG} & LO & NLO &
Data~\cite{HFAG} & LO & NLO 
\\
\hline
$\pi^\pm K^{*0}$ & 
$11.3 \pm 1.0$ & $5.5$ & $6.0^{+2.8}_{-1.5}$ &
$-8.6 \pm  5.6$ & $-3$ & $-1^{+1}_{-0}$
\\
$\pi^0 K^{*\pm}$ & 
$6.9 \pm 2.3$ & $4.0$ & $4.3^{+5.0}_{-2.2}$ &
$\phantom{-}4 \pm 29$ & $-38$ & $-32^{+21}_{-28}$
\\
$\pi^\mp K^{*\pm}$ & 
$9.8\pm 1.1$ & $5.1$ & $6.0^{+6.8}_{-2.6}$ &
$-5 \pm 14$ & $-56$ & $-60^{+32}_{-19}$
\\
$\pi^0 K^{*0}$ & 
$1.7 \pm 0.8$ & $1.5$ & $2.0^{+1.2}_{-0.6}$ &
$-1^{+27}_{-26}$ & $-5$ & $-11^{+7}_{-5}$
\\
\hline 
$\rho^\pm K^{0}$ & 
$ < 48 $ & $3.6$ & $8.7^{+6.8}_{-4.4}$ &
--- & $2$ & $1\pm 1$
\\
$\rho^0 K^{\pm}$ & 
$4.25^{+0.55}_{-0.56}$ & $2.5$ & $5.1^{+4.1}_{-2.8}$ &
$31^{+11}_{-10}$ & $79$ & $71^{+25}_{-35}$
\\
$\rho^\mp K^{\pm}$ & 
$9.9^{+1.6}_{-1.5} $ & $4.7$ & $8.8^{+6.8}_{-4.5}$ &
$17^{+15}_{-16}$ & $83$ & $64^{+24}_{-30}$
\\
$\rho^0 K^{0}$ & 
$5.4 \pm 0.9$ & $2.5$ & $4.8^{+4.3}_{-2.3}$ &
%$-64{}\pm41{}\pm20$ 
$-64\pm 46$
& $7$ & $7^{+8}_{-5}$
\\
\hline 
$\omega K^{\pm}$ & 
$6.9\pm 0.5$ & $2.1$ & $10.6^{+10.4}_{-5.8}$ &
$5\pm 6$ & $82$ & $32^{+15}_{-17}$
\\
$\omega K^{0}$ & 
$4.8\pm 0.6$ & $1.9$ & $9.8^{+8.6}_{-4.9}$ &
$21\pm 19$ & $-4$ & $-3^{+2}_{-4}$
\\
\hline 
$\phi K^\pm$ & 
$8.30\pm 0.65$ & $13.8$ & $7.8^{+5.9}_{-1.8}$ &
$3.4\pm 4.4$ & $-2$ & $1^{+0}_{-1}$
\\
$\phi K^0$ & 
$8.3^{+1.2}_{-1.0}$ & $12.9$ & $7.3^{+5.4}_{-1.6}$ &
$-1\pm 13$ & $0$ & $3^{+1}_{-2}$
\\
\end{tabular}
\end{ruledtabular}
\end{table}
The LO predictions show a hierarchy of the branching ratios: 
Br$(B\to\pi K)$\,$>$\,Br$(B\to\pi K^*)$\,$\gtrsim$\,Br$(B\to\rho(\omega) K)$.  
The hierarchy can na\"{\i}vely be understood by considering penguin
emission amplitudes associated with the effective Wilson coefficients
$a_4$ and $a_6$ shown in Figs.~\ref{fig:bpv}(a) and (b), respectively. 
Because of the destructive (constructive) combination of the $a_4$ and
$a_6$ amplitudes in $B\to\rho(\omega) K$ (in $B\to\pi K$), the branching
ratios for $B\to\rho(\omega) K$ are smaller than those for $B\to \pi K$. 
The smaller $B\to\pi K^*$ ratios are caused by the fact that those modes
have no $a_6$ emission diagram.

Including the NLO corrections, the predictions for the $B\to PV$
branching ratios are in better agreement with the data as shown in
Table~\ref{tab:pv}. The NLO predictions for $B^\pm \to \pi^\pm K^{*0}$,
$\pi^0 K^{*\pm}$, and $\pi^\mp K^{*\pm}$ reach only 2/3 of the measured
values, whereas that for  $B^0 \to \pi^0 K^{*0}$ is in agreement with
the data. Considering the uncertainties of both the theoretical
predictions and the data, the discrepancy is not serious. 
Note that the hierarchy 
Br$(B\to\pi K^*)$\,$\gtrsim$\,Br$(B\to\rho(\omega) K)$ at LO is reversed
into Br$(B\to\pi K^*)$\,$\lesssim$\,Br$(B\to\rho(\omega) K)$ with the
NLO effects. The NLO corrections dramatically enhance the $B\to\rho K$,
$\omega K$ branching ratios. As a result, the NLO predictions for the
$B\to\rho K$ decays are in agreement with the data, while the central
values of the NLO $B\to\omega K$ predictions are higher than the
observed ones. 
The $B\to\omega K$ modes, as well as the $B\to\phi K$ modes, have an
additional $a_5$ amplitude shown in Fig.~\ref{fig:bpv}(c). The vertex
corrections to the $a_5$ amplitude increase (decrease) the 
$B\to \omega K$ ($\phi K$) branching ratios significantly. 
If the hard scale is lifted slightly, the $a_5$ contribution is
moderated. Br$(B\to\omega K)$ and Br$(B\to \phi K)$ then decrease and
increase, respectively, approaching the central values of the data.

The observed patterns of the direct CP asymmetries differ among the 
$B\to \pi K$, $\pi K^*$, and $\rho K$ modes as shown in
Tables~\ref{tab:pik_pipi} and \ref{tab:pv}~\cite{HFAG}.  
As explained in the last section, the penguin amplitude $P'$ in the
$B\to\pi K$ decays is in the second quadrant with a strong phase coming
from the scalar-penguin annihilation diagram in Fig.~\ref{fig:bpv}(d). 
The color-suppressed tree amplitude $C'$ is enhanced by the vertex
corrections, such that $T'+C'$ is almost anti-parallel to $P'$. It leads
to the pattern in Eq.~(\ref{eq:pik}) and a negative asymmetry in the 
$B^0\to\pi^\mp K^\pm$ mode.
In the $B\to\pi K^*$ decays, $P'$ is more inclined to the positive
imaginary axis compared to that in the $B\to\pi K$ decays. 
$T'$ and $P'$ are then more orthogonal to each other, and $T'+C'$ and
$P'$ do not line up. Therefore, PQCD predicts the larger asymmetries
with the pattern~\cite{Li:2006jv} 
\begin{eqnarray}
|A_{\rm CP}(B^\pm\to\pi^0 K^{*\pm})|
&<& 
|A_{\rm CP}(B^0\to\pi^\mp K^{*\pm})| 
\;, 
\label{eq:pikstar}
\end{eqnarray}
as shown in Table~\ref{tab:pv}. 
In the case of the $B\to\rho K$ decays, the real part of $P'$ diminishes
due to the destructive interference between Figs.~\ref{fig:bpv}(a) and
(b). Because of the sign flip of the scalar-penguin annihilation
amplitude, $P'$ is roughly aligned with the negative imaginary
axis. Therefore, the predicted direct CP asymmetries are larger and
positive with~\cite{Li:2006jv}
\begin{eqnarray}
A_{\rm CP}(B^\pm\to\rho^0 K^\pm)
&\approx&
A_{\rm CP}(B^0\to\rho^\mp K^\pm) 
\;.
\label{eq:rhok}
\end{eqnarray}
Thus, NLO PQCD predicts the different patterns of the direct CP
asymmetries among the $B\to \pi K$, $\pi K^*$, and $\rho K$ modes as
shown in Eqs.~(\ref{eq:pik}), (\ref{eq:pikstar}), and (\ref{eq:rhok}).

\section{Mixing-induced CP Asymmetries \label{sec:mixing}}

The mixing-induced CP asymmetries for the penguin-dominated modes are
useful to search for new physics beyond the SM. Their asymmetries are
na\"{\i}vely expected to be identical to those for the tree-dominated
$b\to c\bar cs$ modes, ${\cal S}_{c\bar cs}=\sin(2\phi_1)$. The possible
SM deviation $\Delta {\cal S}_f\equiv {\cal S}_f - \sin(2\phi_1)$ for
the final states $f\,(=\pi^0 K_S,\,\rho^0 K_S,\,\omega K_S,\,\phi K_S)$
can be written as  
\begin{eqnarray}
\Delta {\cal S}_{f} 
\simeq
2 
\left(
\lambda^2\sqrt{\rho^2+\eta^2}\,
-\, {\rm Re}\frac{C'}{P'}
\right)
\cos(2\phi_1)
\sin\phi_3
\;,
\end{eqnarray}
where $\lambda$, $\rho$, and $\eta$ are the CKM parameters. 
For the small $C'$ case, the deviation is small and positive 
($\Delta {\cal S}_f\sim 0.02$).

The NLO PQCD predictions for 
$\Delta {\cal S}_f$~\cite{Li:2005kt,Li:2006jv} are summarized in
Table~\ref{tab:deltaS}, where those in QCDF~\cite{Beneke:2005pu} and in
QCDF plus long-distance (LD) effects~\cite{Cheng:2005bg} are also shown
for comparison. 
\begin{table}
\caption{\label{tab:deltaS} Possible SM deviations of the mixing-induced
 CP asymmetries in PQCD~\cite{Li:2005kt,Li:2006jv}, in
 QCDF~\cite{Beneke:2005pu}, and in QCDF+LD~\cite{Cheng:2005bg}.}
\begin{ruledtabular}
\begin{tabular}{c|cccc}
Mode & Data~\cite{HFAG} & PQCD & QCDF & QCDF$+$LD
\\
\hline
$\pi^0 K_S$ &
$-0.35\pm 0.21$ & %Data
$\phantom{-}0.06{}^{+0.02}_{-0.03}$ & %PQCD
$\phantom{-}0.07{}^{+0.05}_{-0.04}$ & %QCDF
$0.04^{+0.02+0.01}_{-0.03-0.01}$ %QCDF+FSI
\\
$\rho^0 K_S$ &
$-0.48\pm 0.57$ & %Data
$-0.19{}^{+0.10}_{-0.06}$ & %PQCD
$-0.08{}^{+0.08}_{-0.12}$ & %QCDF
$0.04^{+0.09+0.08}_{-0.10-0.11}$ %QCDF+FSI
\\
$\omega K_S$ &
$-0.20\pm 0.24$ & %Data
$\phantom{-}0.15{}^{+0.03}_{-0.06}$ & %PQCD
$\phantom{-}0.13{}^{+0.08}_{-0.08}$ & %QCDF
$0.01^{+0.02+0.02}_{-0.04-0.01}$ %QCDF+FSI
\\
$\phi K_S$ &
$-0.29\pm 0.18$ & %Data
$\phantom{-}0.03{}^{+0.00}_{-0.02}$ & %PQCD
$\phantom{-}0.02{}^{+0.01}_{-0.01}$ & %QCDF
$0.03^{+0.01+0.01}_{-0.04-0.01}$ %QCDF+FSI
\\
\end{tabular}
\end{ruledtabular}
\end{table}
PQCD predicts positive deviations, opposite to the observed ones, except
for the $B^0\to\rho^0 K_S$ case. In the $B\to\rho K$ decays, both $P'$
and $C'$ are almost imaginary and parallel to each other. Therefore,
PQCD predicts the negative $\Delta {\cal S}_{\rho^0 K_S}$. Among the
modes studied here, $B^0\to\phi K_S$ is the cleanest one, since it dose
not involve $C'$. The predictions are basically in agreement with those
in QCDF, but differ from those in QCDF$+$LD, which predicted $\Delta
{\cal S}_{\rho^0 K_S}>0$.

\section{Summary \label{sec:summary}}

In this talk, we have summarized the recent works on exclusive $B$ meson
decays in the PQCD approach, with focus on $B\to\pi K$, $\pi\pi$, and
the penguin-dominated $B\to PV$ modes.

In the NLO PQCD analysis, it has been found that the color-suppressed
tree amplitude is enhanced by the vertex corrections; the direct CP
asymmetries for the $B\to\pi K$ modes have approached the observed
values. The $B^0\to\pi^0\pi^0$ branching ratio is, however, still too
small, compared with the measured value. 
NLO PQCD has also predicted the different patterns of the direct CP
asymmetries among the $B\to\pi K$, $\pi K^*$, and $\rho K$ modes as
shown in Eqs.~(\ref{eq:pik}), (\ref{eq:pikstar}), and (\ref{eq:rhok}). 
Those patterns, if confirmed by data, will support the source of strong
phases from the scalar-penguin annihilation diagrams in the PQCD
approach. 
The predicted mixing-induced CP asymmetries for the penguin-dominated
modes basically show the positive deviations, 
${\cal S}_{\pi^0 K_S,\, \omega K_S,\, \phi K_S} \gtrsim \sin(2\phi_1)$,
except for $S_{\rho^0 K_S} \approx 0.5$.  
Hence, the central values of the current data, 
$\Delta {\cal S}_{\pi^0 K_S,\, \omega K_S,\, \phi K_S}<0$, seem to 
be still puzzling.

\begin{acknowledgments}
I am grateful to the organizers of CKM2006 for a stimulating workshop. 
I would like to thank Hsiang-nan~Li for helpful comments. 
This work was supported by the U.S. Department of Energy under Grant
 No. DE-FG02-90ER40542. 
\end{acknowledgments}


\begin{thebibliography}{99}

\bibitem{HFAG} 
  Heavy Flavor Averaging Group, 
  %``Averages of b-hadron properties at the end of 2005,''
  hep-ex/0603003; updated in
  http://www.slac.stanford.edu/xorg/hfag.
  %%CITATION = HEP-EX/0603003;%%

\bibitem{BBNS}
  M.~Beneke, G.~Buchalla, M.~Neubert, and C.T.~Sachrajda,
  %``{QCD} factorization for B $\to$ pi pi decays: Strong phases and CP 
  %violation in the heavy quark limit,''
  Phys. Rev. Lett. {\bf 83}, 1914 (1999);
  % [arXiv:hep-ph/9905312].
  %%CITATION = HEP-PH 9905312;%%
  %M.~Beneke, G.~Buchalla, M.~Neubert, and C.T.~Sachrajda,
  %``QCD factorization for exclusive, non-leptonic B meson decays:
  %General arguments and the case of heavy-light final states,''
  Nucl. Phys. B {\bf 591}, 313 (2000);
  % [arXiv:hep-ph/0006124].
  %%CITATION = HEP-PH 0006124;%%
  %M.~Beneke, G.~Buchalla, M.~Neubert, and C.T.~Sachrajda,
  %``QCD factorization in B $\to$ pi K, pi pi decays and extraction of
  %Wolfenstein parameters,''
  Nucl. Phys. B {\bf 606}, 245 (2001); 
  % [arXiv:hep-ph/0104110].
  %%CITATION = HEP-PH 0104110;%%
  M.~Beneke and M.~Neubert,
  %``QCD factorization for B $\to$ P P and B $\to$ P V decays,''
  Nucl. Phys. B {\bf 675}, 333 (2003).
  % [arXiv:hep-ph/0308039].
  %%CITATION = HEP-PH 0308039;%%

\bibitem{SCET}
  C.W.~Bauer, S.~Fleming, and M.E.~Luke,
  %``Summing Sudakov logarithms in B $\to$ X/s gamma in effective field 
  %theory,''
  Phys. Rev. D {\bf 63}, 014006 (2001);
  % [arXiv:hep-ph/0005275].
  %%CITATION = HEP-PH 0005275;%%
  C.W.~Bauer, S.~Fleming, D.~Pirjol, and I.W.~Stewart,
  %``An effective field theory for collinear and soft gluons: Heavy to 
  %light decays,''
  Phys. Rev. D {\bf 63}, 114020 (2001);
  % [arXiv:hep-ph/0011336].
  %%CITATION = HEP-PH 0011336;%%
  C.W.~Bauer and I.W.~Stewart,
  %``Invariant operators in collinear effective theory,''
  Phys. Lett. B {\bf 516}, 134 (2001);
  % [arXiv:hep-ph/0107001].
  %%CITATION = HEP-PH 0107001;%%
  C.W.~Bauer, D.~Pirjol, and I.W.~Stewart,
  %``Soft-collinear factorization in effective field theory,''
  Phys. Rev.D {\bf 65}, 054022 (2002).
  % [arXiv:hep-ph/0109045].
  %%CITATION = HEP-PH 0109045;%%

\bibitem{PQCD}
  Y.Y.~Keum, H-n.~Li, and A.I.~Sanda,
  %``Fat penguins and imaginary penguins in perturbative QCD,''
  Phys. Lett. B {\bf 504}, 6 (2001);
  % [arXiv:hep-ph/0004004].
  %%CITATION = PHLTA,B504,6;%%
  %Y.Y.~Keum, H-n.~Li, and A.I.~Sanda,
  %``Penguin enhancement and B --> K pi decays in perturbative QCD,''
  Phys. Rev. D {\bf 63}, 054008 (2001);
  % [arXiv:hep-ph/0004173].
  %%CITATION = PHRVA,D63,054008;%%
  C.D.~Lu, K.~Ukai, and M.Z.~Yang,
  %``Branching ratio and CP violation of B --> pi pi decays in
  %perturbative QCD approach,''
  Phys. Rev. D {\bf 63}, 074009 (2001).
  % [arXiv:hep-ph/0004213].
  %%CITATION = PHRVA,D63,074009;%%

\bibitem{Li:1992nu}
  H-n.~Li and G.~Sterman,
  %``The Perturbative pion form-factor with Sudakov suppression,''
  Nucl. Phys. B {\bf 381}, 129 (1992).
  %%CITATION = NUPHA,B381,129;%%

\bibitem{CL}
  C.H.~Chang and H-n.~Li,
  %``Three-scale factorization theorem and effective field theory:
  %Analysis of nonleptonic heavy meson decays,''
  Phys. Rev.D {\bf 55}, 5577 (1997);
  % [arXiv:hep-ph/9607214].
  %%CITATION = HEP-PH 9607214;%%
  T.W.~Yeh and H-n.~Li,
  %``Factorization theorems, effective field theory, and nonleptonic
  %heavy meson decays,''
  Phys. Rev.D {\bf 56}, 1615 (1997).
  % [arXiv:hep-ph/9701233].
  %%CITATION = HEP-PH 9701233;%%

\bibitem{Li:2005kt}
  H-n.~Li, S.~Mishima, and A.I.~Sanda,
  %``Resolution to the B --> pi K puzzle,''
  Phys. Rev. D {\bf 72}, 114005 (2005). 
  % [arXiv:hep-ph/0508041].
  %%CITATION = PHRVA,D72,114005;%%

\bibitem{Li:2006cv}
  H-n.~Li and S.~Mishima,
  %``Implication of the B $\to$ rho rho data on the B $\to$ pi pi
  %puzzle,'' 
  Phys. Rev. D {\bf 73}, 114014 (2006).
  % [arXiv:hep-ph/0602214].
  %%CITATION = PHRVA,D73,114014;%%

\bibitem{Li:2006jv}
  H-n.~Li and S.~Mishima,
  %``Penguin-dominated B --> P V decays in NLO perturbative QCD,''
  Phys. Rev. D {\bf 74}, 094020 (2006).
  % [arXiv:hep-ph/0608277].
  %%CITATION = PHRVA,D74,094020;%%
	
\bibitem{Brodsky:1982gc}
  S.J.~Brodsky, G.P.~Lepage, and P.B.~Mackenzie,
  %``On The Elimination Of Scale Ambiguities In Perturbative Quantum 
  %Chromodynamics,''
  Phys. Rev. D {\bf 28}, 228 (1983).
  %%CITATION = PHRVA,D28,228;%%

\bibitem{PUZZLE}
  S.~Baek and D.~London,
  %``Is there still a B --> pi K puzzle?,''
  hep-ph/0701181; 
  %%CITATION = HEP-PH/0701181;%%
  R.~Fleischer, S.~Recksiegel, and F.~Schwab,
  %``On puzzles and non-puzzles in B --> pi pi, pi K decays,''
  hep-ph/0702275.
  %%CITATION = HEP-PH/0702275;%%

\bibitem{Mishima:2001ms}
  S.~Mishima,
  %``Understanding the penguin amplitude in B $\to$ Phi K decays,''
  Phys. Lett. B {\bf 521}, 252 (2001); 
  % [arXiv:hep-ph/0107206].
  %%CITATION = HEP-PH 0107206;%%
  C.H.~Chen, Y.Y.~Keum, and H-n.~Li,
  %``Perturbative QCD analysis of B $\to$ Phi K decays,''
  Phys. Rev. D {\bf 64}, 112002 (2001).
  % [arXiv:hep-ph/0107165].
  %%CITATION = HEP-PH 0107165;%%

\bibitem{Beneke:2005pu}
  M.~Beneke,
  %``Corrections to sin(2beta) from CP asymmetries in B0 --> (pi0, rho0,
  %eta, eta', omega, Phi) K(S) decays,''
  Phys. Lett. B {\bf 620}, 143 (2005).
  % [arXiv:hep-ph/0505075].
  %%CITATION = PHLTA,B620,143;%%

\bibitem{Cheng:2005bg}
  H.Y.~Cheng, C.K.~Chua, and A.~Soni,
  %``Effects of final-state interactions on mixing-induced CP violation 
  %in penguin-dominated B decays,''
  Phys. Rev. D {\bf 72}, 014006 (2005). 
  % [arXiv:hep-ph/0502235].
  %%CITATION = PHRVA,D72,014006;%%


\end{thebibliography}
\end{document}